\documentclass[aps,prl,twocolumn,superscriptaddress,showpacs]{revtex4}

\usepackage{color}
\usepackage{graphicx}

\begin{document}

\input epsf.sty
\title{Field-induced ferroelectric and 
commensurate spin state in mutiferroic HoMn$_{2}$O$_{5}$}


\author{H. Kimura}
\email[]{kimura@tagen.tohoku.ac.jp}

\author{Y. Kamada}

\author{Y. Noda}
\affiliation{Institute of Multidisciplinary Research for Advanced Materials, 
Tohoku University, Sendai 980-8577, Japan}

\author{K. Kaneko}

\author{N. Metoki}
\affiliation{Advanced Science Research Center, 
Japan Atomic Energy Agency, Tokai, Ibaraki 319-1195, Japan}

\author{K. Kohn}
\affiliation{Department of Physics, Waseda University, Tokyo 169-8555, Japan}

\date{\today}

\begin{abstract}
Neutron diffraction measurements under high magnetic fields have been performed 
for the multiferroic compound HoMn$_{2}$O$_{5}$. At zero field, high-temperature 
incommensurate magnetic (HT-ICM) -- commensurate magnetic (CM) -- 
low-temperature incommensurate magnetic (LT-ICM) 
orders occur with decreasing temperature, where ferroelectric polarization 
arises only in the CM phase. Upon applying a magnetic field, 
the LT-ICM phase completely disappears and 
the CM phase is induced at the lowest temperature. 
This field-induced CM state is completely associated 
with the field-induced electric polarization in this 
material [Higashiyama {\it et al}., Phys. Rev. B {\bf 72}, 064421 (2005).], strongly 
indicating that the commensurate spin state is essential to the ferroelectricity in 
the multiferroic 
$R$Mn$_{2}$O$_{5}$ system. 
\end{abstract}


\maketitle

Cross-correlations among different extensive (or intensive) variables 
have been one of the most important issues from both scientific and technological 
points of view. Among them, the correlation between magnetic and dielectric properties, 
which is actualized as a magentoelectric (ME) effect, has recently drawn much attention 
and been studied intensively. Although the ME effect, where an electric polarization induced by 
a magnetic field or inversely a magnetization induced by an electric field, has been predicted 
since a long time ago\cite{Curie1894}, the magnitude of the ME effect has been too small 
to reply to an industrial demand thus far. However, it has recently been found that 
$R$MnO$_{3}$ and $R$Mn$_{2}$O$_{5}$ ($R=$ rare earth, Bi, or Y) 
systems show a colossal ME effect\cite{Kimura2003,Hur2004-1}, 
which provides new candidate materials for 
multifunctional devices such as a magnetically controlled 
ferroelectric memory. 
Numerous studies have revealed that these materials show a ``multiferroic'' behavior, 
where magnetic and dielectric orders coexist in almost the same temperature range 
and their order parameters strongly couple 
with each other\cite{Goto2004,Kagomiya2002,Noda2003}. 
However, the microscopic properties of the antiferromagnetism and 
ferroelectricity as well as their relevance 
to the colossal ME effect have not been fully understood yet. To clarify them, 
we have studied the magnetic and dielectric properties of 
the $R$Mn$_{2}$O$_{5}$ system using a neutron diffraction technique. 
Our systematic studies for 
$R$Mn$_{2}$O$_{5}$ compounds with $R=$ Er, Y, Tb and Tm 
have revealed that the incommensurate--commensurate 
phase transition for Mn$^{3+}$ and Mn$^{4+}$ spin structures 
strongly correlates with a ferroelectric phase 
transition\cite{Noda2003,Kagomiya2001,Kobayashi2004-1,Kobayashi2004-2,Kobayashi2004-3,Kobayashi2005,Kobayashi2005-2}. 

Recently, Higashiyama {\it et al}. reported the effects of magnetic field on 
the dielectric properties of the HoMn$_{2}$O$_{5}$ 
compound\cite{Higashiyama2005}. They found that 
the magnetic field along the $b$-axis induced a spontaneous electric polarization, 
indicating that the ferroelectricity in the $R$Mn$_{2}$O$_{5}$ system is magnetically driven. 
However, to clarify the origin of field-induced ferroelectricity, a microscopic 
investigation for magnetic properties as well as for the lattice distortion 
that arises from the ferroelectricity is indispensable. In this work, we thus 
performed a neutron diffraction study under a magnetic field for a 
HoMn$_{2}$O$_{5}$ single crystal 
to establish the magnetic phase diagram for the spin structure and its relevance to 
the dielectric phase diagram. 
Our measurements reveal that the incommensurate--commensurate phase transition on 
the spin structure occurs when the magnetic field is applied along the $b$-axis, showing that 
the commensurate spin state is essential to the ferroelectricity in HoMn$_{2}$O$_{5}$. 

A single crystal of HoMn$_{2}$O$_{5}$ was grown by a 
PbO--PbF$_{2}$ flux method\cite{Wanklyn1972}. The size of the crystal was about 100~mm$^{3}$. 
Neutron diffraction measurements were performed using the thermal neutron triple-axis spectrometer 
TAS-2 installed at JRR-3M in the Japan Atomic Energy Agency. 
The sample was mounted on the ($h\ 0\ l$) 
scattering plane with a vertical-field superconducting magnet, which results in applying a 
field along the $b$-axis perfectly. The incident and final energies of neutrons were fixed at 
14.3~meV using a pyrolytic graphite (PG) (002) monochromator and an analyzer. A PG filter was 
inserted in front of the sample to reduce higher-order contamination. The experimental 
configuration was 15'-80'-80'-80'. In the field cooling process, we applied 
the magnetic field at 60~K, far above $T_{\rm N}$. We also 
performed magnetic structure analysis at the commensurate magnetic phase 
under zero magnetic field using an identical HoMn$_{2}$O$_{5}$ single crystal. 
Details of the structure analysis involving 
the results for YMn$_{2}$O$_{5}$ and ErMn$_{2}$O$_{5}$ 
will be published elsewhere\cite{Kimura2006}. 

\begin{figure}[tbp]
\centerline{\epsfxsize=2.8in\epsfbox{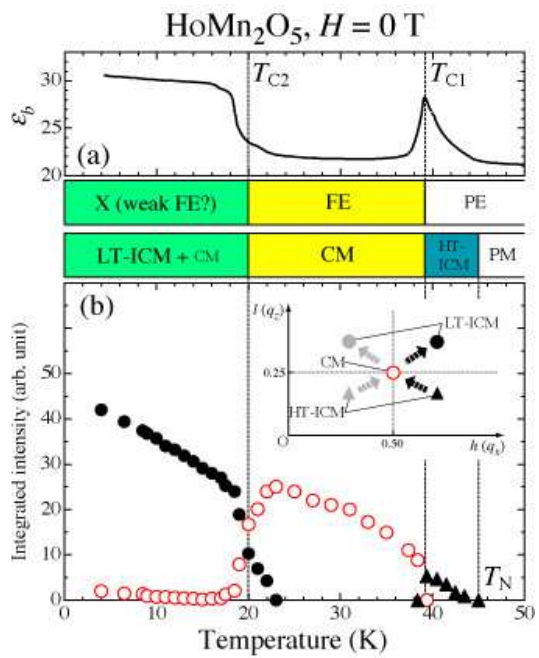}}
\caption{
(Color online) 
Temperature dependences of (a) dielectric constant along $b$-axis and 
(b) integrated intensities of LT-ICM (filled circles), CM (open circles), and 
HT-ICM (filled triangles) Bragg peaks, taken at $H=0$~T. 
The middle part of this figure shows a schematic phase diagram for 
dielectric and magnetic properties. Positions of magnetic 
peaks (propagation wave vectors $q_{x}$, $q_{z}$) 
in LT-ICM, CM, and HT-ICM phases are depicted in 
the inset of Fig.~\ref{fig1}(b). 
}
\label{fig1}
\end{figure}

HoMn$_{2}$O$_{5}$ has an orthorhombic structure with $Pbam$ symmetry at room temperature, 
where edge-sharing Mn$^{4+}$O$_{6}$ octahedra align along 
the $c$-axis and pairs of Mn$^{3+}$O$_{5}$ pyramids link the Mn$^{4+}$O$_{6}$ chains 
in the $ab$-plane\cite{Abrahams1967}. 
The 4$f$-moment of Ho$^{3+}$, and the 3$d$-moment of Mn$^{4+}$ and Mn$^{3+}$ 
ions are responsible for the characteristic magnetism in this system. 
In zero magnetic field, we have obtained magnetic and dielectric properties 
analogous to those of other $R$Mn$_{2}$O$_{5}$ compounds reported 
previously\cite{Kagomiya2001,Kobayashi2004-1,Kobayashi2004-2,Kobayashi2004-3,Kobayashi2005,Kobayashi2005-2}. 
Figure~\ref{fig1}(a) shows dielectric constant as a function of temperature, 
where two phase transitions are seen apparently at $\sim 39$~K ($\equiv T_{\rm C1}$) and 
$\sim 20$~K ($\equiv T_{\rm C2}$). It was reported that the spontaneous polarization along 
the $b$-axis grows up below $T_{\rm C1}$, whereas the polarization becomes weaker or almost 
vanishes below $T_{\rm C2}$\cite{Higashiyama2005}. In this paper, we 
respectively call the intermediate $T$ phase between $T_{\rm C1}$ and $T_{\rm C2}$, and 
the lowest $T$ phase below $T_{\rm C2}$, 
the FE and X phases, as defined in Ref.~\cite{Higashiyama2005}. 
The temperature dependence of the intensity of 
magnetic Bragg peaks is shown in Fig.~\ref{fig1}(b). Filled and open symbols 
indicate the intensity observed at incommensurate and commensurate {\bf Q} positions, respectively. 
As seen in Fig.~\ref{fig1}(b), a high-temperature incommensurate magnetic (HT-ICM) phase 
appears (closed triangles) below $T_{\rm N}$ and disappears at $T_{\rm C1}$, 
where the commensurate magnetic (CM) 
phase (open circles) and the ferroelectric (FE) phase arises concurrently. 
On further cooling, the CM phase vanishes and 
a low-temperature incommensurate magnetic (LT-ICM) phase (closed circles) emerges 
around $T_{\rm C2}$, where the spontaneous polarization 
becomes almost zero\cite{Higashiyama2005}. 
The inset in Fig.~\ref{fig1}(b) 
shows the {\bf Q} position of the magnetic Bragg peak denoted with a 
propagation wave vector ${\bf q}=(q_{x}\ 0\ q_{z})$ in three different magnetic phases. 
Note that the minor CM phase persists even below $T_{\rm C2}$, which coexists 
with the major LT-ICM phase. Both the dielectric and magnetic phase diagrams 
as functions of temperature are schematically shown in the middle panel of Fig.~\ref{fig1}. 

\begin{figure}[tbp]
\centerline{\epsfxsize=3.6in\epsfbox{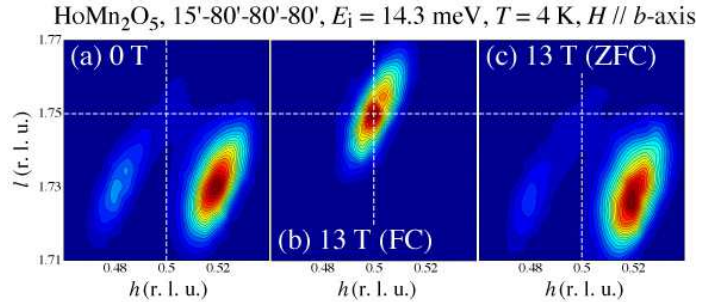}}
\caption{
(Color online) 
Intensity maps of LT-ICM and CM Bragg peaks around $(\frac{1}{2}\ 0\ 2$-$\frac{1}{4})$ 
in $(h\ 0\ l)$ zone at $T=4$~K. Data was taken under 
(a) $H=0$~T, (b) $H=13$~T in field cooling (FC), and 
(c) $H=13$~T in zero-field cooling (ZFC). 
The magnetic field was applied along the $b$-axis.
}
\label{fig2}
\end{figure}

On applying the magnetic field along the $b$-axis up to 13~T, the spin structure 
at the lowest temperature markedly changes. 
Figure~\ref{fig2} shows contour maps of LT-ICM and CM Bragg intensities 
around ${\bf Q}=(\frac{1}{2}\ 0\ 2$-$\frac{1}{4})$ at $T=4$~K, taken under zero field and $H=13$~T. 
These maps correspond to slices along the $(h\ 0\ l)$ reciprocal plane. At zero field, as seen in 
Fig.~\ref{fig2}(a), two peaks are found at the incommensurate position described 
as the propagation vector of ${\bf q}\sim(\frac{1}{2}$$\pm$0.019\ 0\ $\frac{1}{4}$+0.022). 
However, when the magnetic field of 13~T is applied on the field cooling (FC) process, 
the two incommensurate peaks completely disappear and 
a single peak appears exactly at the commensurate position, shown in 
Fig.~\ref{fig2}(b). Figure~\ref{fig2}(c) shows the data taken at 13~T with the zero-field cooling 
(ZFC) process, where the incommensurate peaks still persist. 
The difference between the peak distribution in the FC process and that in the ZFC process 
means that there is a large hysteresis at the incommensurate--commensurate phase 
transition, suggesting the first-order nature 
of the phase transition. The first-order nature 
is also seen in the dielectric phase transition between the FE and 
X phases\cite{Higashiyama2005}. This one-to-one correspondence 
between the magnetic and dielectric properties clearly shows 
that the FE phase is magnetically induced. 

\begin{figure}[tbp]
\centerline{\epsfxsize=2.6in\epsfbox{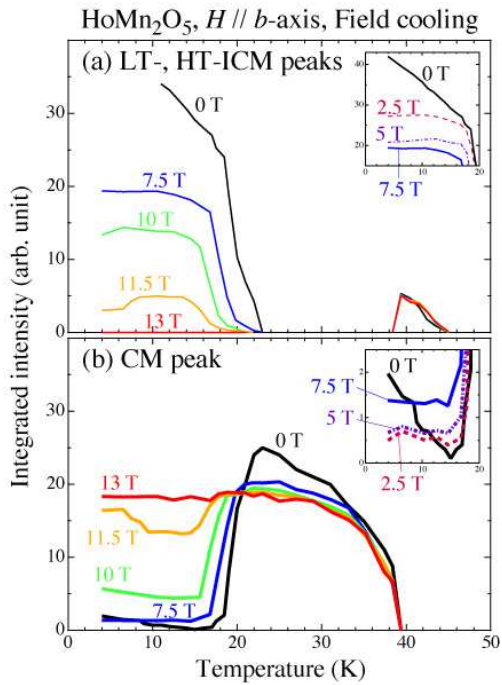}}
\caption{
(Color online) 
Temperature dependence of integrated intensity for (a) LT-ICM, HT-ICM and 
(b) CM Bragg peaks, taken under $H=0$~T and $7.5\leq H\leq 13$~T. Insets show 
enlarged views of the data below $T=20$~K and $0\leq H\leq 7.5$~T.
}
\label{fig3}
\end{figure}

The detailed field dependences of order parameters for the LT-ICM, HT-ICM, and CM phases 
are summarized in Fig.~\ref{fig3}. 
All the data were taken in the FC process up to $H=13$~T. 
Figure~\ref{fig3}(a) shows the temperature dependence of 
intensity for LT- and HT-ICM Bragg peaks. The behavior of the HT-ICM signal 
at around 40~K, 
involving the appearance and disappearance temperatures and the intensity variation, 
is completely independent of magnetic field. In contrast, the appearance temperature of 
the LT-ICM peak and its intensity decrease with increasing field. 
As for the CM signal shown in Fig.~\ref{fig3}(b), the appearance temperature 
($T_{\rm C1}$) and the evolution 
of the order parameter above $\sim 20$~K do not change significantly, whereas 
the intensity of the signal below $\sim 20$~K is enhanced with increasing magnetic field. 
The field dependence for LT-ICM and CM signals becomes marked in the region 
between 10~T and 13~T, where both phases coexist and the fraction of intensity 
in each phase changes with field. This behavior indicates 
that incommensurate--commensurate magnetic phase transition has a strong 
first-order nature, which is consistent with 
the large hysteretic behavior seen in Fig.~\ref{fig2}. The insets of Fig.~\ref{fig3} 
show the detailed temperature dependence below $T=20$~K 
under magnetic fields between 0~T and 7.5~T. 
At zero field, not only the major LT-ICM but also 
the minor CM phases evolve with decreasing temperature. However, above 
$H=2.5$~T, the temperature evolutions of both phases are almost saturated. 
Although the details are not clarified yet, this behavior might indicate the field 
response related with a Ho$^{3+}$ moment 
induced by the ordering of Mn$^{3+}$ and Mn$^{4+}$ moments. 
Indeed, the magnetic structure 
analysis under zero field using the identical single crystal shows that 
Ho$^{3+}$ moments of up to $\sim 1.3 \mu_{\rm B}$ are induced 
even at 25~K (CM phase)\cite{Kimura2006}. 

\begin{figure}[tbp]
\centerline{\epsfxsize=2.8in\epsfbox{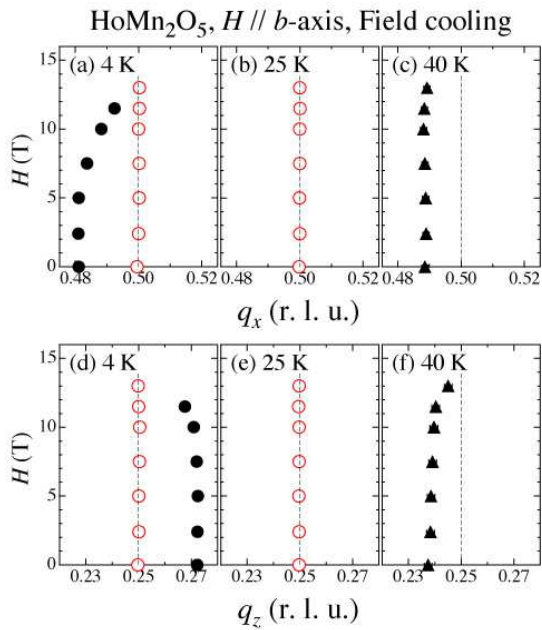}}
\caption{
(Color online) 
Field dependence of propagation wave vector ${\bf q}=(q_{x}\ 0\ q_{z})$ taken at 
$T=4$~K, 25~K, and 40~K. (a)-(c), and (d)-(f) correspond to the results for 
$q_{x}$ and $q_{z}$ components, respectively. Filled circles, open circles, and filled triangles 
denote the data from LT-ICM, CM, and HT-ICM peaks, respectively. 
}
\label{fig4}
\end{figure}

A magnetic propagation wave vector described by ${\bf q}=(q_{x}\ 0\ q_{z})$ 
as a function of magnetic field and temperature was measured. 
Figure~\ref{fig4} shows the field dependence of ${\bf q}$ taken at 
$T=4$~K, 25~K, and 40~K, corresponding to the LT-ICM, CM, and HT-ICM 
phases under zero field, respectively. The results for $q_{x}$ and 
$q_{z}$ components appear on 
the upper panels (Figs.~\ref{fig4}(a)--(c)) and 
the lower panels (Figs.~\ref{fig4}(d)--(f)), respectively. 
It is clearly seen that as magnetic field increases, 
the ${\bf q}$ vectors for the CM and HT-ICM phases 
(open circles and filled triangles) are almost field independent, whereas 
those for the LT-ICM phase (filled circles) approach a commensurate position, 
indicating that the LT-ICM phase 
becomes unstable due to the development of the CM phase. 
It should be noted that the $q_{z}$ component at 40~K also comes close to 
the 1/4 position at $H=13$~T. This might indicate that 
a one-dimensionally modulated ICM order is induced by the magnetic field, 
the phase of which was observed in ErMn$_{2}$O$_{5}$\cite{Kobayashi2004-1} 
and YMn$_{2}$O$_{5}$\cite{Kobayashi2004-2} at zero field. 

\begin{figure}[tbp]
\centerline{\epsfxsize=3.1in\epsfbox{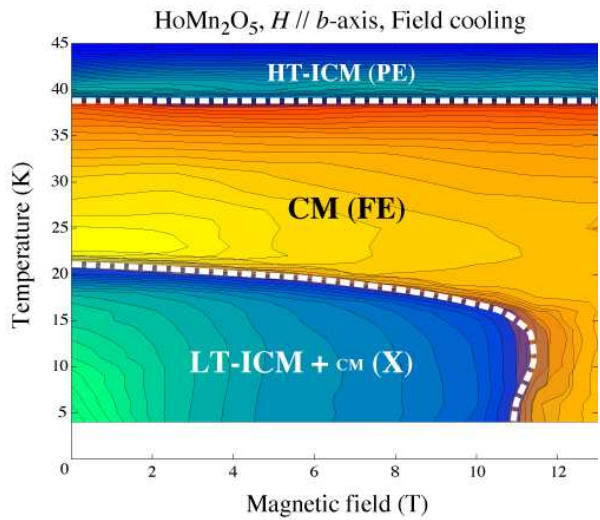}}
\caption{
(Color online) 
$H-T$ phase diagram for ICM and CM states in HoMn$_{2}$O$_{5}$, 
shown as the contour map of intensity for ICM and CM magnetic signals. 
The dielectric phase diagram 
taken by Higashiyama {\it et al}.(ref.~\cite{Higashiyama2005}) is superposed 
in this figure. 
}
\label{fig5}
\end{figure}

To overview the correlation between the field response of the microscopic 
magnetic property and that of the dielectric property, we mapped out the 
magnetic field ($H$) -- temperature ($T$) phase diagram 
for LT-ICM, CM, and HT-ICM states. Figure~\ref{fig5} shows the 
contour map of the intensities of the magnetic signals as functions of temperature 
and magnetic field only in the FC process. 
Boundaries between each phase are roughly indicated by white dashed lines. 
Distributions for each phase are expressed by a gradation sequence. 
The obtained phase diagram apparently shows that 
HT-ICM is completely field independent, whereas 
the LT-ICM phase rapidly disappears and the CM phase is induced 
around $H\sim 11$~T. Comparing our $H$--$T$ phase diagram with 
the dielectric phase diagram\cite{Higashiyama2005}, it is quite interesting that 
the LT-ICM and CM phases in the magnetic property 
perfectly correspond to the null or weak ferroelectric (X) phase and 
the robust ferroelectric (FE) phase, respectively. These experimental facts 
evidently establish that the commensurate spin state 
is indispensable to the ferroelectricity, and thus the dielectric property 
in this system has a magnetic origin. 

Our results have shown that the magnetic field strongly affects the magnetism below 
$\sim 20$~K ($T_{\rm C2}$ at zero field), whereas the magnetism above $\sim 20$~K 
is almost field independent. This strongly suggests that the 4$f$-moment 
of rare-earth atoms, of which 
amplitude becomes larger as temperature decreases, significantly contributes 
to the field responses in the magnetic and dielectric properties. 
Actually, as many studies have reported, a large ME effect has been found only in 
the lower-temperature region (LT-ICM phase at zero field)\cite{Higashiyama2005,Hur2004-2}. 
Recent symmetric analysis of the magnetic 
structure for (Tb,Ho,Dy)Mn$_{2}$O$_{5}$ has shown that the magnetic interaction 
in Mn$^{3+}$ and Mn$^{4+}$ sublattices is geometrically 
frustrated\cite{Chapon2004,Blake2005}. Such magnetic frustration 
might induce a competing magnetic ground state, that can be easily tuned by 
the $f-d$ spin exchange interaction between $R^{3+}$ and Mn$^{3+, 4+}$ ions. 
Our magnetic structure analysis in 
the CM phase of (Y,Er,Ho)Mn$_{2}$O$_{5}$ single crystals also clarified that 
the magnetic structure of Mn ions is common regardless of 
the existence of the $f$ moment\cite{Noda2003,Kimura2006}. Combining 
these results with the present observation of the field-induced CM state in 
HoMn$_{2}$O$_{5}$, it is strongly suggested that 
the commensurate magnetic ground state in Mn ions is indispensable to 
the ferroelectricity in $R$Mn$_{2}$O$_{5}$ compounds. 

Microscopic mechanisms of the magnetically driven ferroelectricity have 
been theoretically proposed with various frameworks such as exchange 
striction\cite{Goodenough1955} and 
Dzyaloshinskii-Moriya (D--M) interaction\cite{Katsura2005,Sergienko2006}. 
However, all the predictions do not require the commensurate spin state. 
The detailed magnetic and crystal structure analyses in both 
LT-ICM and CM phases are essential for 
completing the microscopic correlation between 
the magnetic and dielectric properties of the multiferroic $R$Mn$_{2}$O$_{5}$ system. 

In summary, we have found the magnetic-field-induced commensurate spin state in 
HoMn$_{2}$O$_{5}$. The field evolution of the commensurate ground state 
completely traces the development of the ferroelectric phase. 
Incommensurate--commensurate phase transition has a first-order nature, which is 
also consistent with a first-order ferroelectric phase transition. This study 
has microscopically shown that the commensurate spin state is 
essential to the ferroelectricity in the $R$Mn$_{2}$O$_{5}$ system. 

We thank M. Matsuda and Y. Shimojo for technical assistance in operating the 
superconducting magnet. 
This work was supported by a Grant-In-Aid for Scientific Research (B), 
No. 16340096, from the Japanese Ministry of
Education, Culture, Sports, Science and Technology. 
\begin{acknowledgments}
\end{acknowledgments}


\end{document}